# Code Arcades: 3d Visualization of Classes, Dependencies and Software Metrics

Anthony Savidis[1,2] and Christos Vasilopoulos[2]

[1] Computer Science Department, University of Crete, Greece
[2] Alpha Omega Zed SA, Research and Development Department, Greece
`savidis@csd.uoc.gr, christos.vasilopoulos@alphaomegazed.com`

**Abstract.** Software visualization seeks to represent software artifacts graphically in two or three dimensions, with the goal of enhancing comprehension, analysis, maintenance, and evolution of the source code. In this context, visualizations employ graphical forms such as dependency structures, treemaps, or timelines that incorporate repository histories. These visualizations allow software engineers to identify structural patterns, detect complexity hotspots, and infer system behaviors that are difficult to perceive directly from source text. By adopting metaphor-based approaches, visualization tools provide macroscopic overviews while enabling focused inspection of specific program elements, thus offering an accessible means of understanding large-scale systems. The contribution of our work lies in three areas. First, we introduce a configurable grouping mechanism that supports flexible organization of code elements based on arbitrary relationships. Second, we combine fine-grained and coarse-grained software metrics to provide a multi-level perspective on system properties. Third, we present an interactive visualization engine that allows developers to dynamically adjust rendering attributes. Collectively, these advances provide a more adaptable and insightful approach to source code comprehension.

## 1  Introduction

Software visualization is concerned with the graphical representation, in 2d or 3d, of the various software-related artifacts, with the purpose of supporting comprehension, analysis, error detection, maintenance and evolution. In the context of source code, common visual forms relate to dependency graphs, tree maps and update timelines. For the latter, the source code repositories with the commit history are directly inspected, thus not only the final source code may be subject to such visualization-driven analysis. The primary goal is to enable software engineers detect and comprehend structural patterns, complexity hotspots, and other non-functional characteristics, otherwise difficult or impossible to perceive directly in the original textual form.

Effectively, source code visualization tools provide macroscopic graphical views by relying on well-chosen visual metaphors, while focusing on particular source code elements to provide an easily understandable structural view for various implementation aspects. In particular, the adoption of spatial metaphors like code cities (Wettel & Lanza 2008, Wettet et al., 2011) and landscape views (Langelier et al., 2005) aim to reveal large variations of complexity, size and other metrics with directly evident peaks or spikes, thus enabling software engineers to directly comprehend and then interactively explore likely issues within large-scale source code bases. Overall, combined with the main software engineering disciplines, visualization plays a key role in supporting source code comprehension, management and supervision at a macroscopic perspective, as outlined under Fig. 1.

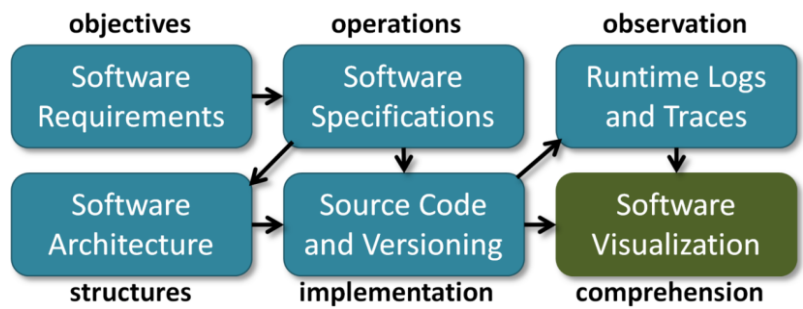

**Fig. 1.** The critical complementary role of software visualizations in the entire software lifecycle for system comprehension and supervision.

### 1.1 Contribution

Our primary contribution concerns three key features when compared to earlier tools for metaphor-based software visualization such as code cities: (i) configurable grouping mechanism enabling place within the same block classes not merely related to packages but to any other ad-hoc grouping mechanism such as architecture recovery relationships (Savidis & Savaki, 2022); (ii) combination of fine-grained metrics such as function size, total formal arguments, total class members and call sites with coarse-grained metrics like method access (public / private), inheritors (is-a), components (part-of), and deployments (objects as arguments); and (iii) configurable visuals enabling interactively changing most attributes of our rendering engine.

## 2 Related Work

Early efforts in metaphor-based software visualization explored spatial and landscape metaphors to help developers grasp abstract code structures. For instance, in (Chuah & Eick, 1998) a landscape metaphor is introduced to model aspects such as project risk, cost, and scheduling through a terrain-like layout. Similarly, in (Jeffery, 2019) a broader overview of the city metaphor is provided, in which software artifacts are depicted as buildings or streets, facilitating comprehension of structural evolution and developer activity.

Reviews have synthesized how different metaphors support various comprehension tasks. In (Bacher et al., 2017), a comprehensive survey of the code-map metaphor, analyzing numerous publications representing various tools. Their study uncovered primarily qualitative reports on usability and effectiveness and also highlighted the lack of rigorous quantitative validation. This gap underscores a recurring issue in metaphor-based visualizations: usability is often assumed rather than empirically tested.

Building upon metaphorical foundations, recent tools have introduced novel visualizations grounded in semantics and structural relationships. For example, in (Wettel & Lanza, 2008), the CodeCity project extended the city metaphor to object-oriented systems, where classes are visualized as buildings and packages as city districts, enabling intuitive exploration of large systems. The latter formed the basis of our idea on Code Arcades. Beyond cities, in (Lungu et al., 2013), the Software Galaxy approach is introduced, applying an astronomy metaphor where topic modeling clusters semantically related code into a galaxy-like visualization, enhancing semantic comprehension. Similarly, SysMap (Teyseyre et al., 2021) employed multiple architectural metaphors—such as solar system, city, and park—to depict software evolution across versions in 3D, aiming to reduce manual effort in understanding system change over time.

In immersive and interactive contexts, metaphor-based visualization takes on new dimensions. In (Steinmetz et al., 2019), VR- and AR-based software visualizations are investigated, exploring spatial metaphors embedded in virtual worlds—ranging from cosmic to extended city metaphors—and allowing immersive navigation of code structure. These prototypes suggest that immersive metaphorical environments can support program understanding and visual programming, while also raising questions about psychological and ergonomic factors.

Overall, there has been important earlier work regarding metaphor-driven approaches, such as landscapes, cities, galaxies, and maps, each leveraging familiar spatial analogies to aid comprehension. However, common limitations persist: many works emphasize conceptual or qualitative merits rather than empirical validation, while skipping aspects that are critical in software engineering like dependencies, metrics and the ability to easily configure them interactively. Our work seeks to advance this with emphasis on developer utility and usability.

## 3    Functionality

We provide an overview of the key rendering and interaction features of the Code Arcades system, emphasizing facilities that go beyond existing tools. As it will be explained later on, when discussing implementation details, many of these features rely upon added-value metadata and semantic information that is produced during preprocessing time, where a massive parsing, analysis and global symbol table generations stage takes place on the entire source code repositories. As in code cities, our primary visualization entities are classes and packages, with classes rendered as buildings and packages as areas or blocks (districts in code cities terminology).

### 3.1 Visualizing Packages and Classes

In our approach there are two ways to denote packages: (i) as C++ namespaces that are directly derived from the original source code; and (ii) as likely architecture components computed following an architecture recovery process like the one introduced in (Savidis & Savaki, 2022). Both are important, since the first designates the initially defined package split, while the latter reflects the actual architecture decomposition as derived from a deep dependency analysis over the source code. Also, the classes of a single package are visually grouped together by topologically arranging them in a left-to-right and front-to-back fashion with a number of alternative and configurable ordering criteria such as: number of classes, source code size (in lines), total commits, number of contributors and summative age (using heuristic to judge how recent a package is based on the last updates of its contained files).

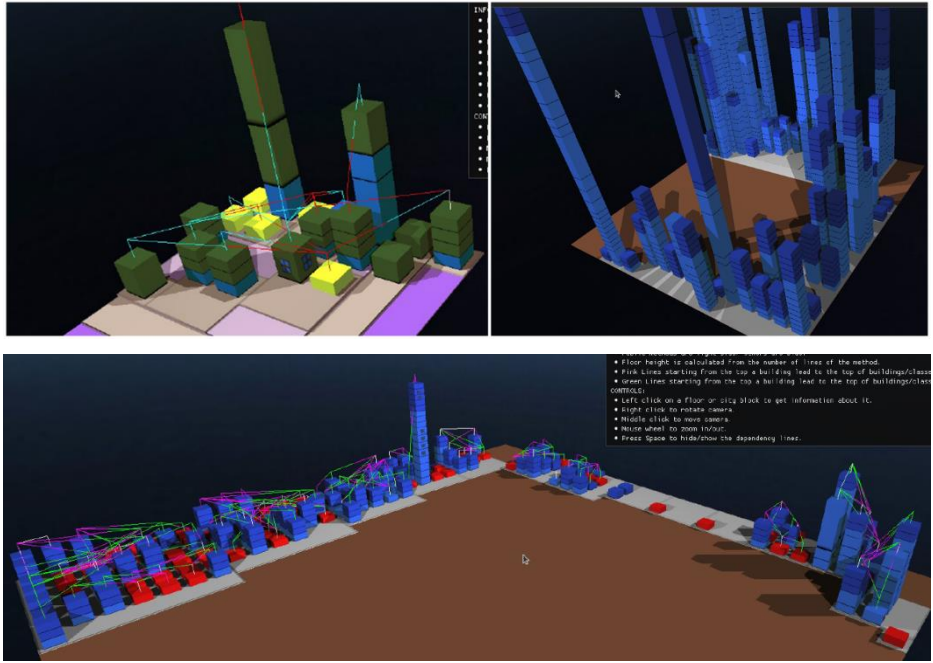

**Fig. 2.** Top: packages as areas and classes as buildings and skyscrapers - floor coloring encodes member-access qualifiers; Bottom: the Code Arcades project visualizing itself.

Every building corresponds to a class and every method to a single floor, while windows designate method arguments. As shown under Fig. 2 (top-right part), the number of methods in a class may be so vast that it leads to typical skyscraper structures, while in other cases thin classes are depicted with just a few floors. This imbalance is directly perceivable with such visualizations, while together with dependency density (incoming or outgoing links) can give important information for class criticality with-in a package. As also mentioned, configurable color encoding is enabled for various semantic aspects enabling immediate visual mapping:

- Classes with no methods are colored in red (PODs – plain old data types in C++)
- Public / private methods are drawn in light / normal blue color
- Dependencies are colored with the half starting and half ending parts in different colors to designate referent (pink) and referrer (cyan) - we found arrows too crowded in 3d)

### 3.2 Visualizing Dependencies

We mentioned earlier the way dependencies are visualized as half-split bicolored lines separate the color of the source (dependent) and target classes (depended on).

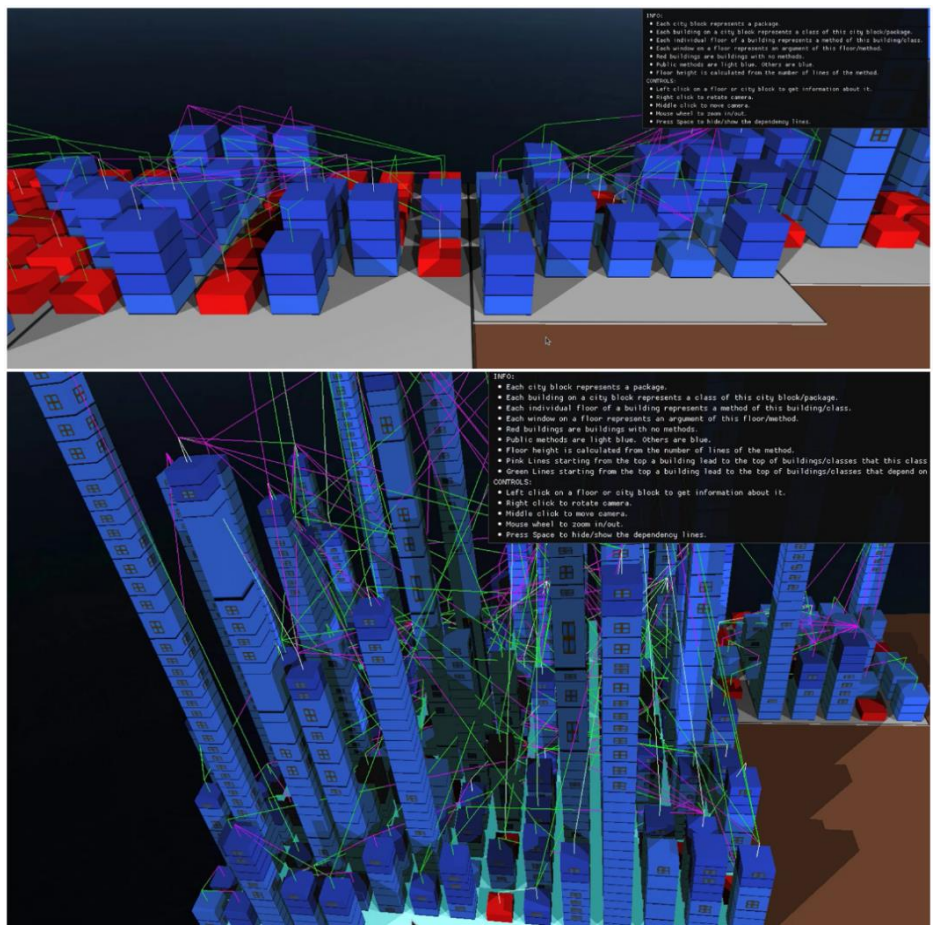

**Fig. 3.** Dependencies drawn between classes for various semantic relationships; for instance, coloring in magenta depicts inheritance (is-a)

The types of dependencies tracked in our system is offered by the backbone tool for Architecture Recovery elaborated under (Savidis & Savaki, 2022). In particular, all dependencies among classes, captured by the adopted architecture recovery tool, are fully exported to Code Arcades, resulting in the cross-building connection lines explained earlier. More specifically, these dependencies concern the following semantics cases:

− A class base of B (A is-a B)
− A object member of B (A part-of B)
− A object used by B (A pointer or reference deployed in B via any of these cases: field, argument to a method, locally used object in a method)
− A class indirectly instrumenting implementation of B (A is used in a template instantiation that is deployed by B)

It should be noted that compared to earlier approaches like cope cities, our support for dependency rendering offers various advantages, as it allows developers to macroscopically recognize related "bad smells" (Mantyla et al., 2004). For instance, dense interdependencies across distinct packages may indicate non-modular design, since packages aim to imply modules which require loose coupling. Additionally, superdense connectivity between a group of classes demonstrates that the original class split was inappropriate, leading to arbitrary and dependency intensive links, thus implying a likely required class-merge action.

### 3.3 Visualizing Software Metrics

Software metrics have been introduced as a way to assess various aspects of source code through quantitative characteristics, in order to judge key properties such as code quality, readability, extensibility, maintainability and complexity.

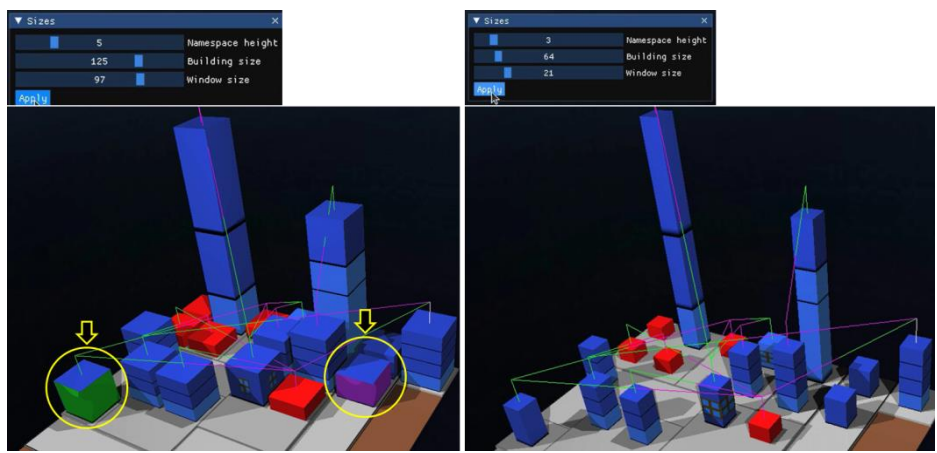

**Fig. 4.** Interactive configuration on metric visuals shown for: classes (buildings), namespaces (blocks) and formal arguments (windows); code smells also modify visual parameters, as shown for the two classes (yellow arrows and circles) whose texture and color is affected

They provide objective insights which enable software engineers make calculated decisions, identify likely issues or risks in the software lifecycle and overall ensure that the code aligns with quality standards. Effectively, as mentioned in (Kan, 2002), early integration of metrics into the software lifecycle is very crucial. In Code Arcades, one of our objectives has been the direct depiction of metrics in the visualization enabling developers immediately identify potential malpractices or issues.

In this context, we have built a predicate system that computes potential code smells, based on source code information and analysis, and then allow smells to affect the visual parameters of the related constructs (classes or methods), such as coloring, dimensions, illumination, material and texture. An example of such configuration during visualization is provided in Fig. 4.

## 4 Architecture

The Code Arcades tool is built on top of a number of other tools we have developed earlier, as outlined within Fig. 5. The *backbone* is implement on top of the Clang Tooling, which in the LLVM/Clang ecosystem refers to all libraries and infrastructure for building tools based on Clang's AST, source parsing, and refactoring capabilities. In particular we introduced custom symbol analysis and repeated parsing, by accessing the compilation database for the particular current project, to populate a global symbol table for the entire software system under inspection.

Then, the second tool in the process is a large-scale system for architecture recovery, explained under (Savidis & Savaki, 2022), which performs dependency analysis, produces a global class graph, and then eventually infers likely architecture components by applying a number of graph clustering algorithms. The output of this tool is an alternative grouping of classes into modules, compared to namespaces, that is also provided as input to the Code Arcades tool.

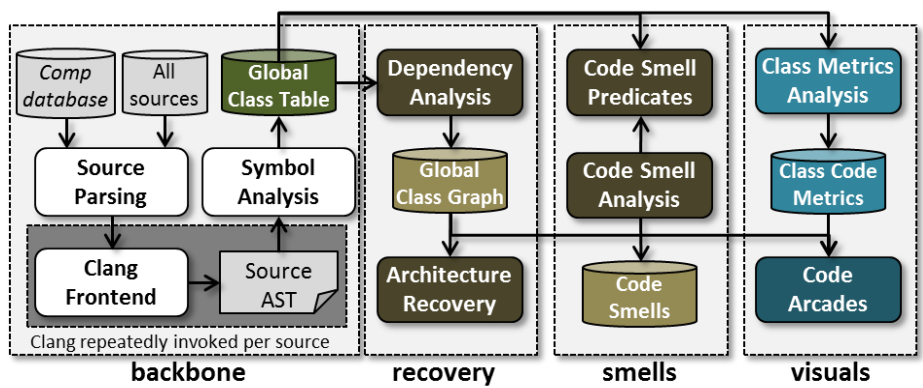

**Fig. 5.** The software architecture of the Code Arcades tool relies on the outcomes of earlier developed components.

Then, the code smells detector is an extensible tool, comprising a repository of code smell predicates. This tool accesses the entire global symbol table, performing predicate evaluation and producing a catalogue of the detected smells, tagged with class and method information from the original symbol table for cross-reference.

The Code Arcades is currently the last tool in the chain, since it accesses all outputs from the previous tools to provide visual insights to the code structure, related metrics and properties.

## 5   Implementation

The implementation of the Code Arcades tool is entirely written in C++ and has been developed on top of the following subsystems and third-party open tools:

— The rendering system is built on top of the "Ogre 3d" open-source software library (https://www.ogre3d.org/)
— The interactive configuration GUI uses the compact "Dear Im Gui" widget library for Open GL (https://www.dearimgui.com/)
— The backbone, as mentioned, relies on Clang Tooling (https://clang.llvm.org/)
— Initial prototypes before implementing the renderer were made by exporting OBJ files for Blender 3d (https://www.blender.org/) and testing with various illumination coloring and texturing setups
— Textures have been designed using Gimp (https://www.gimp.org/), the GNU image manipulation tool

## 6   Conclusions

In conclusion, Code Arcades advances the field of software visualization by providing a novel 3D environment where classes, methods, dependencies and software metrics are rendered in a way that is both configurable and interactive. By integrating fine-grained and coarse-grained metrics with metaphor-based visual representations, the tool enables developers to detect complexity hotspots, design flaws, and code smells at a glance.

Such macroscopic views enable ease of comprehension and effective supervision of source code aspects in ways that are impossible directly in the source text form. The incorporation of flexible grouping mechanisms - whether based on namespaces or likely components from an architecture recovery process - further enhances its adaptability to different software engineering contexts. These features collectively address several limitations of earlier approaches, such as rigid metaphors, limited configurability, and insufficient emphasis on dependencies, ultimately making large-scale systems more comprehensible and manageable.

Looking ahead, the modular design of Code Arcades provides a foundation for further exploration into immersive, empirical, and collaborative uses of software visualization. Future extensions may include integration with version-control history for evolution analysis, VR/AR support for immersive inspection, and empirical usability studies to validate its effectiveness in real-world large-scale industrial projects. More broadly, Code Arcades contributes to a growing body of work that frames visualization not just as a supplement to textual source analysis, but as a core practice in modern software engineering. By aligning visual metaphors with quantitative metrics and developer-centric interaction, the tool demonstrates how visualization can play a critical role in improving comprehension, quality assurance, and long-term maintainability of complex software systems.